\documentclass[amssymb,aps,superscriptaddress,twocolumn,floats,showpacs]{revtex4-2}

\usepackage{color,graphicx,pstricks,float}
\usepackage{natbib}

\usepackage{fancybox}
\usepackage{epsfig}
\usepackage{tabularx}
\usepackage{inputenc}
\usepackage{amsmath}
\usepackage{amssymb}
\usepackage{textcomp}
\usepackage{gensymb}
\usepackage{array,xcolor}
\usepackage{amsbsy}
\usepackage{braket}
\usepackage{color,graphicx,pstricks,float}
\definecolor{newgreen}{rgb}{0,0.4,0.2}

\usepackage{url,hyperref}
\hypersetup{
	colorlinks = true,
	linkcolor = blue,
	filecolor = blue,      
	urlcolor = blue,
	citecolor = blue,
} 

\begin{document}

\title{First Order Topological Phase Transitions and Disorder Induced Majorana Modes in Interacting Fermion Chains}

\author{Shruti Agarwal}
\affiliation{Department of Physical Sciences,
Indian Institute of Science Education and Research Mohali, Sector 81, S.A.S. Nagar, Manauli PO 140306, India}
\author{Shreekant Gawande}
\affiliation{Department of Physical Sciences,
Indian Institute of Science Education and Research Mohali, Sector 81, S.A.S. Nagar, Manauli PO 140306, India}
\author{Satoshi Nishimoto}
\affiliation{Department of Physics, Technical University Dresden, 01069 Dresden} 
\affiliation{Institute for Theoretical Solid State Physics, IFW Dresden, 01069 Dresden, Germany}
\author{Jeroen van den Brink}
\affiliation{Department of Physics, Technical University Dresden, 01069 Dresden} 
\affiliation{Institute for Theoretical Solid State Physics, IFW Dresden, 01069 Dresden, Germany}
\author{Sanjeev Kumar}
\affiliation{Department of Physical Sciences,
Indian Institute of Science Education and Research Mohali, Sector 81, S.A.S. Nagar, Manauli PO 140306, India}
\affiliation{Institute for Theoretical Solid State Physics, IFW Dresden, 01069 Dresden, Germany}

\begin{abstract}
Using a combination of the mean-field Bogoliubov deGennes (BdG) approach and the Density Matrix Renormalization Group (DMRG) method, we discover first order topological transitions between topological superconducting and trivial insulating phases in a sawtooth lattice of inter-site attractive fermions. Topological characterization of different phases is achieved in terms of winding numbers, Majorana edge modes and entanglement spectra. By studying the effect of disorder on the first order topological phase transitions, we establish the disorder-induced topological phase coexistence as a mechanism for generating a finite density of Majorana particles.
\end{abstract}
\date{\today}

\maketitle

{\it Introduction:--}
Discovery of topological insulators marked the beginning of a paradigm shift in our approach to understand ordering phenomena in condensed matter physics \cite{Kane2005e, Murakami2004, Sheng2006, Fu2007, Groth2009, Hasan2010b}. Topological phases, since then, have become a recurring theme of research in many fields of quantum as well as classical physics \cite{Albert2015, Socolar2017, Goldman2016, Kane2013, Yang2015, Lu2014a, Lee2018}. Search for new topological phases is driven, not merely by our curiosity to achieve a fundamental understanding but also by the potential for their applications. Topological superconductors are of particular importance as these can host Majorana particles that are considered to be building blocks of quantum computers \cite{Flensberg2021, Qi2011a, Beenakker2013a, Alicea2011, Das2012, Nayak2008a}. Topological phases and phase transitions in non-interacting models can be comprehensively understood in terms of the symmetries of the Hamiltonian \cite{Altland1997, Ryu2010, Cornfeld2021, Song2020a}. On the other hand, general understanding of topological phases in interacting systems is a challenging theoretical problem \cite{Guo2011, Lutchyn2011, Tang2012, Manmana2012, Ezawa2017}.

Phase separation (PS) is a ubiquitous phenomena displayed by electronic systems \cite{Kremer1994, Furrer1994, Hofmann2020}. While the inability of an electronic system to exist in a uniform density state appears to be an undesirable feature, PS has proven to be of critical importance in understanding certain aspects of the physics of strongly correlated electron systems \cite{Johnston1994, Hizhnyakov1994}. Invoking electronic PS is for instance essential for theories of colossal magnetoresistance effect in manganites \cite{Dagotto2003}. Also some of the theories applicable in the low-doping regime of the famous cuprate superconductors rely on electronic PS \cite{DiCastro1994, Kugel2008, DeMello2003}. 
To the best of our knowledge, electronic PS between a topological and a trivial phase -- henceforth referred to as topological PS (TPS) -- has not been discussed in literature. TPS has general implications that are independent of the detailed parameter choices of the model.

In this Letter, we report the discovery of first order topological phase transitions in a lattice model of attractively interacting spinless fermions. Using a combination of DMRG and mean-field BdG methods, we find that the model hosts topological superconductor (TSC) and charge-modulated insulator (CMI) phases. Some of the topological and trivial phases are separated by first order boundaries, leading to TPS.
We characterize the topological phases with the help of the winding numbers, edge modes and the entanglement spectra. 
We explicitly demonstrate that in the presence of quenched disorder the TPS leads to phase coexistence with Majorana modes residing at the boundaries of TSC regions. To the best of our knowledge, this is the first mechanism for generating a finite density of Majorana particles in a system of interacting fermions.

\noindent
{\it Spinless fermions on sawtooth lattice:--}
\begin{figure}[H]
    \centering
    \includegraphics[width= 0.98\columnwidth,angle=0,clip=true]{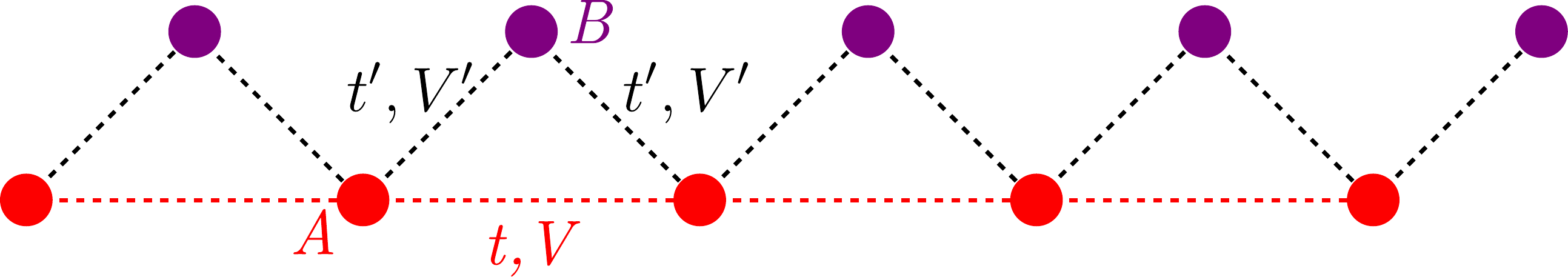}
    \caption{Schematic representation of the sawtooth lattice model. $t$, $t'$ indicate the hopping amplitudes and $V$, $V'$ the attractive interactions between electrons on neighboring sites. Inequivalent sites are labeled A and B.}
    \label{fig:model}
\end{figure}
Let us consider spinless fermions with attractive interactions residing on a sawtooth lattice described by the Hamiltonian, 
\begin{eqnarray}
	H & = & - t \sum_i (c^\dagger_{i, A} c^{}_{i+1, A} + {\textrm H.c.}) -t' \sum_{i} (c^{\dagger}_{i, A} c^{}_{i, B} + {\textrm H.c.}) \nonumber \\ 
	& & - t' \sum_{i} (c^{\dagger}_{i, B} c^{}_{i+1, A} + {\textrm H.c.}) - V \sum_i \hat{n}_{i, A} \hat{n}_{i+1,A} \nonumber \\
	& & -V' \sum_i (\hat{n}_{i, A} \hat{n}_{i,B} + \hat{n}_{i, B} \hat{n}_{i+1,A}) - \mu \sum_{i,s} \hat{n}_{i,s}.
	\label{eq:Ham1}
\end{eqnarray}
Here, $c_{i, s} (c_{i, s}^\dagger$) annihilates (creates) an electron at Bravais site $i$ and sublattice $s \in \{ A, B \}$, and $\hat{n}_{i,s}$ is the corresponding number operator. Hopping amplitudes are denoted by $t$ and $t'$, and the corresponding attractive interaction strengths by $V$ and $V'$ (see FIG. \ref{fig:model}). We set $t'=1$ as the basic energy scale, and $\mu$ denotes the chemical potential. Note that this elementary interacting model Hamiltonian reduces to a monoatomic Kitaev chain supporting p-wave superconductivity for $t = V = 0$, and the non-interacting limit displays a flat band dispersion for $t'/t = \sqrt{2}$.

\begin{figure}[t!]
    \centering
    \includegraphics[width= 0.98\columnwidth,angle=0,clip=true]{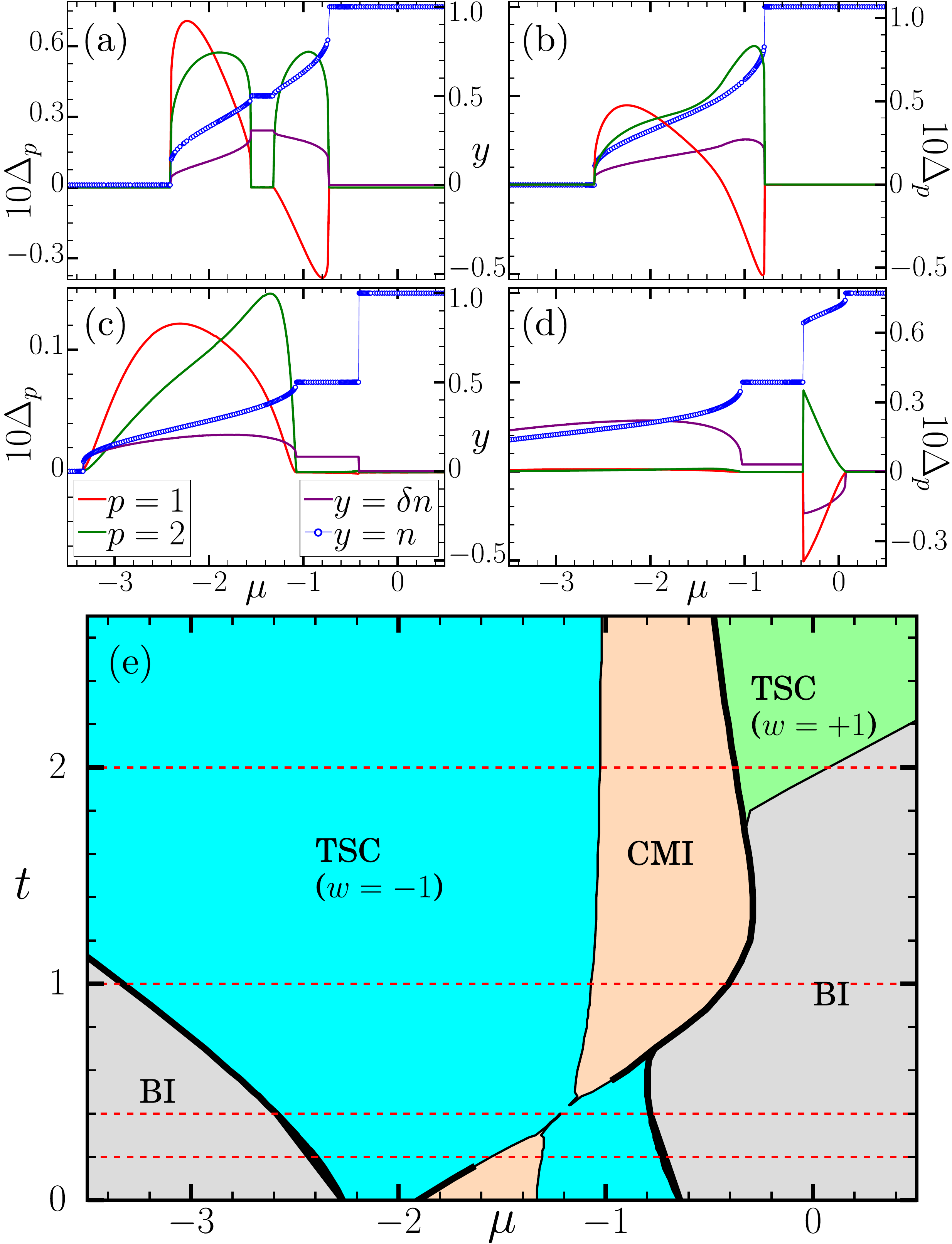}
    \caption{Zero-temperature variations of $\Delta_1$, $\Delta_2$, $\delta n$ and $n$ with chemical potential for $V = V' = 1$ and for, (a) $t = 0.2$, (b) $t = 0.4$, (c) $t = 1.0$, (d) $t = 2.0$. (e) Ground state phase diagram in the $t-\mu$ plane constructed from the data similar to that shown in panels (a)-(d). Horizontal dashed lines indicate the scans for which the data in panels (a)-(d) is displayed.}
    \label{fig:1}
\end{figure}

{\it Mean Field Phase Diagram:--}
In order to obtain the mean-field ground states of the Hamiltonian Eq. (\ref{eq:Ham1}), we decouple the interaction term in pairing and density channels \cite{SM,Zhu2016a,1998,PhysRevB.63.020505}. This leads to the effective non-interacting Hamiltonian, 
\begin{eqnarray}
 H& = &\frac{1}{2}\sum_k \Psi_k^\dagger H_k \Psi_k,
\label{eq:hk0}
\end{eqnarray}
\noindent
with,
\begin{eqnarray}
\Psi_k &=&  \begin{bmatrix} c^\dagger_{k,A} & c^\dagger_{k,B}  & c_{-k,A} & c_{-k,B} \end{bmatrix}^\dagger,
H_k = \begin{bmatrix} \beta_k & \alpha_k \\ \alpha^\dagger_k & -\beta^T_{-k} \end{bmatrix}, \nonumber \\ 
\beta_k &=& \begin{bmatrix} -\Tilde{\mu}_A-2t\cos k & ~ ~ ~ ~ -t'(1+e^{-\mathrm{i}k}) \\ -t'(1+e^{\mathrm{i}k}) & ~ ~ ~ ~  -\Tilde{\mu}_B \end{bmatrix}, \nonumber \\
\alpha_k &=& \begin{bmatrix} -2\mathrm{i} V \Delta_1 \sin k & ~ ~ ~ ~  -V'\Delta_2(1-e^{-\mathrm{i} k}) \\ V'\Delta_2(1-e^{\mathrm{i} k}) & ~ ~ ~ ~ 0 \end{bmatrix}, 
\label{eq:hk}
\end{eqnarray}
\noindent
$\Tilde{\mu}_A = \mu+2n_AV+2n_BV'$ and  $\Tilde{\mu}_B = \mu + 2n_A V'$. The mean-field parameters, $n_{A/B} = \braket{\hat{n}_{i,A/B}}$, $\Delta_1 = \braket{c^{}_{i+1,A} c^{}_{i,A}}$ and $\Delta_2 = \braket{c^{}_{i+1,A} c^{}_{i,B}} = \braket{c^{}_{i,B} c^{}_{i,A}}$ are self-consistently determined \cite{SM}. FIG. \ref{fig:1}(a)-(d) displays the variation of different mean-field parameters with chemical potential. The density per site, $n = (n_A+n_B)/2$, varies from $0$ to $1$, with the end points representing trivial band insulator (BI) phases corresponding to empty or fully filled bands. The pair expectation values, $\Delta_1$ and $\Delta_2$, are found to be finite at all non-trivial densities except when there is a plateau at $n=0.5$ (see FIG. \ref{fig:1}(a), (c), (d)). The difference in average densities at two sublattices, $\delta n = n_B - n_A$, remains finite in the entire region. The finite $\delta n$ is related to the inequivalence of the two sublattices in the non-interacting tight-binding model. Finite values of $\Delta_1$ and $\Delta_2$ identify a superconducting state, within the mean-field description. The $n=0.5$ plateau regions are characterized by a gap in the single-particle density of states (DOS), together with vanishing pairing amplitudes and finite $\delta n$. Therefore, we label these states as charge modulated insulators (CMI). We summarize the variations of mean-field parameters obtained for different values of $t$ in terms of a ground state phase diagram in FIG. \ref{fig:1}(e). The topological superconductor (TSC) phases are further labeled as $w = \pm 1$, where $w$ denotes the winding number.  
Note that the total density displays discontinuities as a function of $\mu$. This is a direct indicator of a first-order phase transition and associated electronic PS in the model. The discontinuities are also present in other mean-field parameters. While most of the discontinuities seem to involve the trivial BI as one of the states, we also find a robust jump in $n(\mu)$ between the TSC and CMI (see FIG. \ref{fig:1}(d)).
The PS locations in $\mu$ become extended lines in the phase diagram and are indicated with thick black lines in FIG. \ref{fig:1}(e).

\begin{figure}
    \centering
    \includegraphics[width = 0.98\columnwidth]{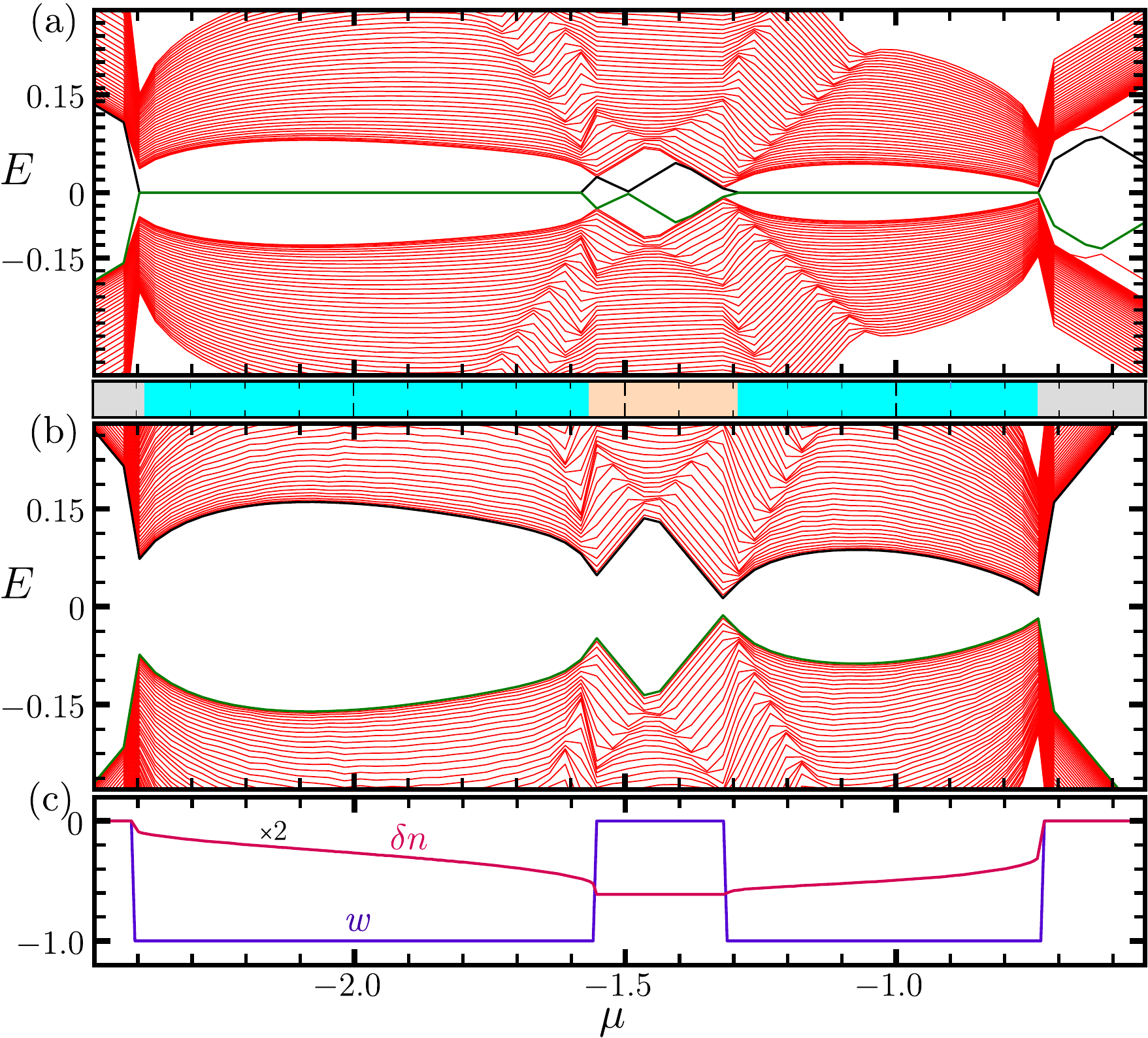}
    \caption{Evolution of the BdG spectrum with $\mu$ for, (a) OBC and (b) PBC. The corresponding variations in the winding number $w$ and the charge modulation $\delta n$ are shown in (c). The values of the parameters used for these calculations are: $t' =V = V'= 1 $, $t = 0.2$}
    \label{fig:edge_state}
\end{figure}

{\it Topological characterization:--}
The low energy quasiparticle spectra is shown as a function of chemical potential for both, periodic boundary conditions (PBC) and open boundary conditions (OBC) in FIG. \ref{fig:edge_state}. The OBC spectra displays states at exactly zero energy over the $\mu$-range corresponding to the superconducting state. We have explicitly checked that these states are Majorana zero energy modes (MZMs) localized on the edges. In the corresponding PBC spectra, the naive expectation is that the bulk gap must close at the transition point between topologically trivial and nontrivial phases. While the gap seems to reduce, it does not close at the transition point. This seemingly unusual feature is easy to understand if we note that change in $\mu$ does not represent a continuous evolution of the effective non-interacting Hamiltonian in parameter space. This is because the self-consistent values of various mean-fields are also parameters of the Hamiltonian which change discontinuously across the transitions. Note that within the CMI phase, two states seem to be crossing the gap in the OBC spectra. We have confirmed that these are trivial edge states that appear due to the difference in effective on-site potentials on the boundary sites. 

Given that we are dealing with a one-dimensional system, winding number turns out to be the most natural choice of the invariant that characterizes topologically distinct phases. Using the standard approach to unitary transform the Hamiltonian into an off-diagonal form \cite{Gao2015}, we find,
\begin{eqnarray}
    UH_kU^{-1} &= & \begin{bmatrix}
       0  & \beta_k + \alpha_k \\
       (\beta_{k} + \alpha_{k})^\dagger  & 0
    \end{bmatrix}.
    \label{odh}
\end{eqnarray}
\noindent
The winding number is then defined as \cite{Gao2015},
\begin{eqnarray}
    w = \frac{1}{2\pi \mathrm{i}} \int_{}^{} dk  \frac{1}{\det [\beta_k + \alpha_k]} \frac{d}{dk}\det[\beta_k + \alpha_k] 
\end{eqnarray}
$|w|$ counts how many times $\det[\beta_{k} + \alpha_{k}]$ winds around the origin. 
We find that $w$ takes values 0, $\pm$1 for the Hamiltonian, where $w$ = 0 corresponds to topologically trivial phases and a non-zero winding number corresponds to topologically non-trivial phases.
Phases with non-zero $w$ are labeled as TSC in the phase diagrams. A representative plot displaying the variation of $w$ with $\mu$ is shown in FIG. \ref{fig:edge_state}(c). 
We find that the sign of $w$ is perfectly correlated with the sign of $\delta n$, and has an interesting interpretation in terms of the SO(3) theory of intertwined charge and superconducting orders \cite{SM}. 

\begin{figure}
	\centering
\includegraphics[width= 0.98\columnwidth,angle=0,clip=true]{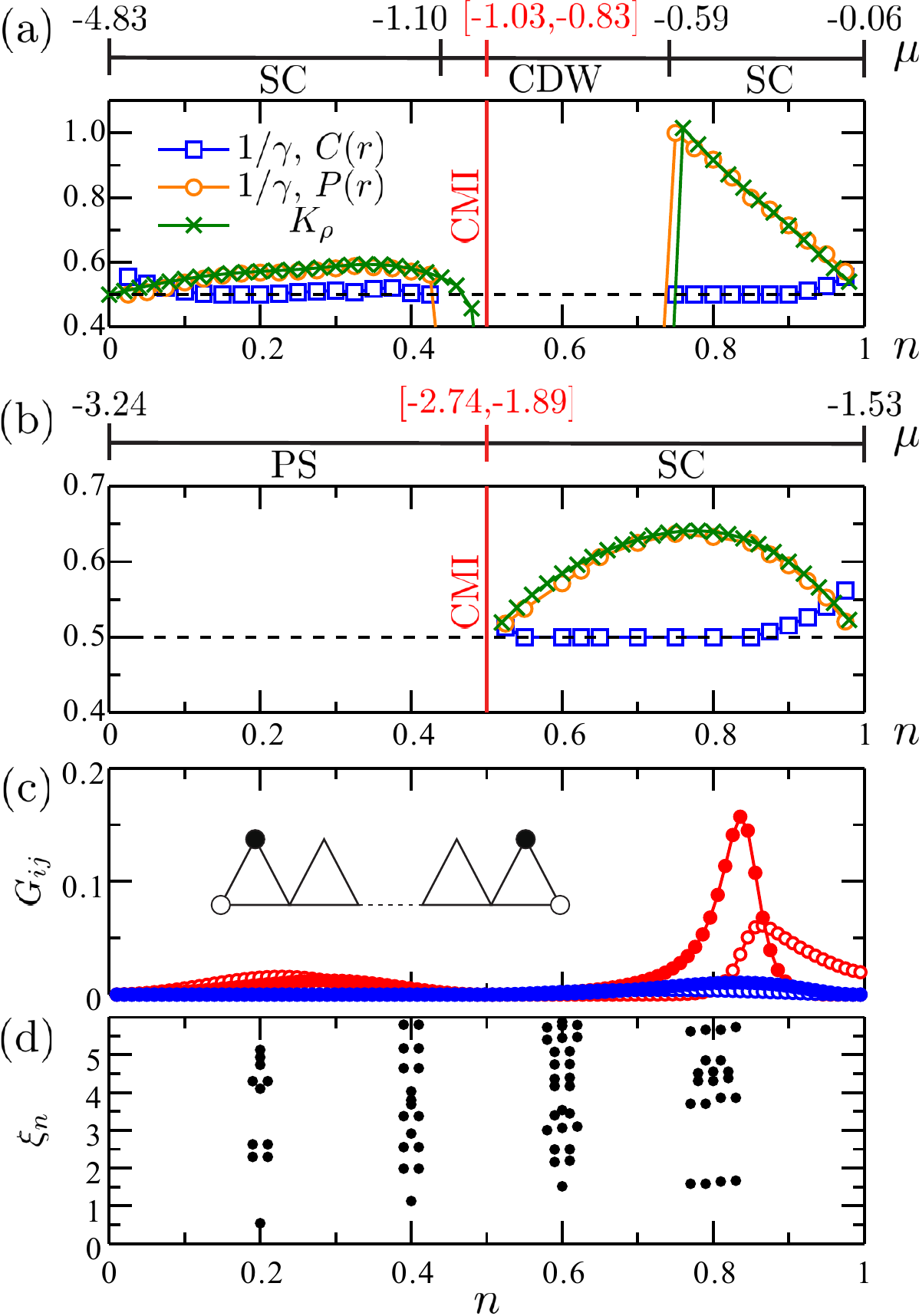}
	\caption{(a) Inverse exponent of the density-density and pair-pair correlation functions as well as the Tomonaga-Luttinger parameter as a function of density per site for $t=2.0$ and $t'=V=V'=1.0$. The corresponding phase diagram is also shown on the top. (b) Similar plot to (a) for $t=t'=V'=1.0$ and $V=3.0$. (c) Edge component of the single-particle correlation function for $t=2.0$, $t'=V=V'=1.0$ (red) and $t=t'=V'=1.0$, $V=3.0$ (blue) with $N=201$ open cluster. The open and filled circles denote the correlations between edged A sites and between edged B sites, respectively. (d) Entanglement spectra as a function of density per site for $t=2.0$, $t'=V=V'=1.0$ with $N=60$ periodic cluster.
}
	\label{fig:DMRG}
\end{figure}
\begin{figure*}
    \centering
    \includegraphics[width = \linewidth]{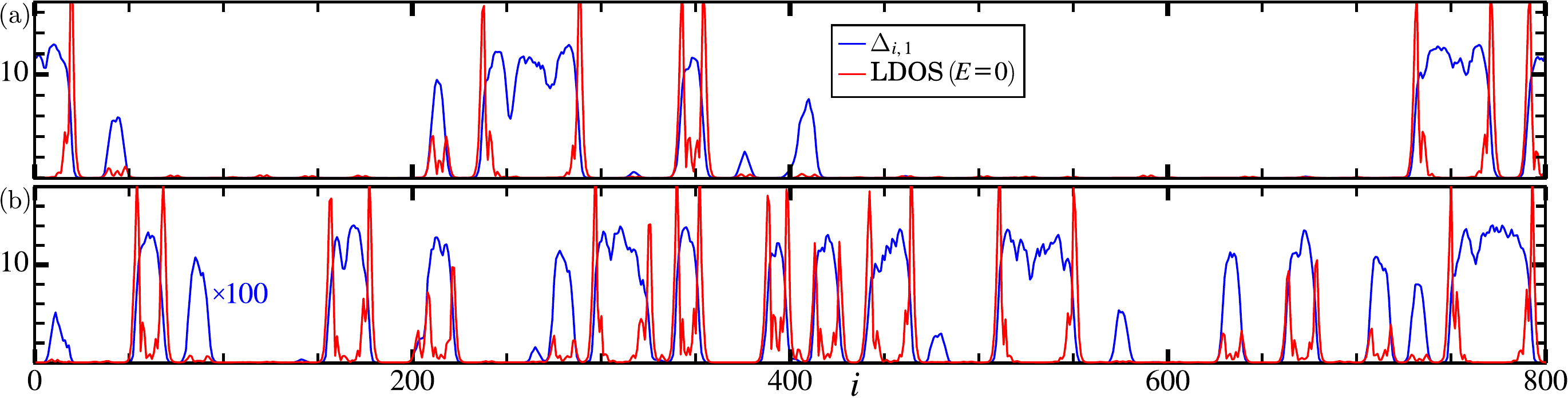}
    \caption{$\Delta_{i,1}$ and zero-energy LDOS as a function of site index $i$ for, (a) $D=0.4$, $n=0.46$ and (b) $D=0.8$, $n=0.435$. The other parameters values are $V = 3$, $V' = 1$, $t = 1$, $\mu = -2.93$. Note that the $\Delta_{i,1}$ values are scaled up in order to emphasize correlations with the LDOS data.}
    \label{fig:phase_sep}
\end{figure*}
{\it DMRG results:--}
In order to check the stability of results beyond mean-field approximation, we perform DMRG calculations for some typical parameter sets. First, the density-density $\langle\hat{n}_{i,s}\hat{n}_{j,s'}\rangle$ and pair-pair $\langle P_{ii'ss'}^\dagger P_{jj'tt'}\rangle$, where $P_{ijss'}=c_{i, s}^\dagger c_{j, s'}^\dagger$, correlation functions are calculated to obtain the ground state phase diagram. The electronic state is characterized by the dominant correlation function with slower decay as a function of distance. The inverse exponent of the power-law decayed correlation functions ($\propto 1/|i-j|^\gamma$) is plotted as a function of density per site in FIG. \ref{fig:DMRG}(a)-(b). The resultant phase diagrams, as a function of $\mu$, are also shown. Note that the SC correlation function decays exponentially in the CDW phase. The phase diagrams are further confirmed by Tomonaga-Luttinger parameter $K_\rho$, which can be obtained accurately via the derivative of charge structure factor at $q=0$ as $K_\rho=\frac{1}{2} \lim_{q \to 0} \langle n(q) n(-q) \rangle$ with $q=2\pi/N$ and $n(q)=\sum_l (e^{-iql}c_{l,A}^\dagger c_{l,A}+e^{-iq(l+1/2)}c_{l,B}^\dagger c_{l,B}$) \cite{Ejima2005}. The dominant SC correlation is indicated when $K_\rho>0.5$. Next, to test the topological nature of the SC phases, the single-particle correlation function $G_{ij}=i\langle\lambda_{i,s}\bar{\lambda}_{j,s´}\rangle$, where $\lambda_{i,s}=c_{i, s}+c_{i, s}^\dagger$ and $\bar{\lambda}_{i,s}=(c_{i, s}-c_{i, s}^\dagger)/i$ are Majorana fermion operators, between two ends of open cluster is calculated \cite{Miao2018}. In FIG. \ref{fig:DMRG}(c) we find a significant enhancement of $G_{ij}$ in the SC phase around $n=0.8$ for $t=2.0$ and $t'=V=V'=1.0$. This indicates that the SC phase is topologically nontrivial. The topological nature is further confirmed by the degeneracy of entanglement spectra $\xi_n(X=Y)$ using the Schmidt decomposition of the ground state $\ket{\psi}$ as $\ket{\psi}=\sum_{\lambda}e^{-\xi_{\lambda}(X)/2} \ket{\lambda}_{X}\ket{\lambda}_{Y}$, where $\ket{\lambda}_{X}$ and $\ket{\lambda}_{Y}$ are the orthonormal basis for the subregions $X$ and $Y$, respectively (see FIG. \ref{fig:DMRG}(d)). Thus, two of the key features of the mean-field results, the TSC phases and PS, are also confirmed by the DMRG analysis. In contrast to the mean-field results, the DMRG calculations also find a CDW phase and a trivial SC phase (see FIG. \ref{fig:DMRG}(a)-(b)).
This may be because the instability to trivial PS, as a particle condensation into a self-bound system of concentration, is missing in the mean-field analysis. It is known that the dominance of $s$-wave conventional SC appears near the trivial PS \cite{Dagotto1994}. Accordingly, a competition between TSC and SC could occur in the DMRG calculations \cite{SM}.

{\it Topological phase coexistence and Majorana modes:--}
Having discussed the phase diagrams, the important features associated with some of the phases and their stability beyond mean-field, we now focus on the effect of quenched disorder on TPS. We add an on-site disorder term, $\sum_i \epsilon_i (\hat{n}_{i,A} + \hat{n}_{i,B})$, to the Hamiltonian Eq. (\ref{eq:Ham1}), where $\epsilon_i \in (-D/2, D/2)$ are random variables drawn from a uniform distribution. Since the translational invariance is lost in the presence of disorder, we are forced to perform calculations in real space. The BdG mean-field Hamiltonian in the presence of disorder can be written as,
\begin{eqnarray}
     H & = &\frac{1}{2} \Psi_r^\dagger H_{r} \Psi_r + E_0,
     \label{eq:real-H}
\end{eqnarray}
\noindent
where, the Nambu spinor $\Psi_r$ contains $4N$ components obtained by repeating the index $i$ from $1$ to $N$ in,
\begin{eqnarray}
     \Psi_r& = & \begin{bmatrix} c^\dagger_{i,A} & c^\dagger_{i,B}  & c_{i,A} & c_{i,B} & ...\end{bmatrix}^\dagger 
\end{eqnarray}
\noindent
$H_r$ is a Hermitian $4N \times 4N$ matrix with the non-zero diagonal and off-diagonal $4 \times 4$ blocks given by, 
\begin{eqnarray} 
h_{ii} &=& \begin{bmatrix} \epsilon_i - \Tilde{\mu}_{i,A} & -t' & 0 & -V'\Delta_{i,2} \\      -t' & \epsilon_i - \Tilde{\mu}_{i,B} & V' \Delta_{i,2} & 0  \\ 
0 & V' \Delta^*_{i,2} & -(\epsilon_i - \Tilde{\mu}_{i,A}) & t' \\
-V'\Delta^*_{i,2} & 0 & t' & -(\epsilon_i - \Tilde{\mu}_{i,B}) 
\end{bmatrix} \nonumber \\
h_{i,i+1} &=& \begin{bmatrix} -t & 0 & -V\Delta_{i,1} & 0 \\                   -t' & 0 & -V' \Delta_{i,3} & 0  \\ 
V \Delta^*_{i,1} & 0 & t & 0 \\
V'\Delta^*_{i,3} & 0 & t' & 0 
\end{bmatrix} = h^*_{i+1,i},
\end{eqnarray}
\noindent 
where, $\Tilde{\mu}_{i,A} = \mu + V(n_{i-1,A} + n_{i+1,A}) + V'(n_{i-1,B}+n_{i,B})$ and $\Tilde{\mu}_{i,B} = \mu + V'(n_{i,A}+n_{i+1,A})$.
The above Hamiltonian is diagonalized via real-space Bogoliubov transformations \cite{Zhu2016a,1998,PhysRevB.63.020505}, and the site-dependent quantum expectation values of density and pairing operators are iteratively computed to obtain self-consistent solutions. We demonstrate the key result by selecting a range of $\mu$ values that cover the phase separation between TSC and CMI for $V=3$, $V'=1$ and $t=1$. We display the spatial dependence of the pairing amplitude $\Delta_{i,1}$, along with the local DOS at zero energy, in FIG. \ref{fig:phase_sep}. Regions with finite pairing amplitudes coexist with regions with zero pairing amplitudes. The number of segments corresponding to the two types of regions increase upon increasing disorder strength (compare (a) and (b) in FIG. \ref{fig:phase_sep}). Most importantly, we find that at the edges of SC regions there exist sharp peaks in zero-energy LDOS. These peaks can only arise from the Majorana zero modes that reside on the edges of all SC segments. It is also clear from the plots that if the length of the SC segment is small then the two edge modes can hybridize. Our calculations explicitly show that TPS provides an interesting route for generating a finite density of Majorana zero modes. 

{\it Conclusion:--}
By investigating a lattice model of attractively interacting fermions, we have unveiled a mechanism for generating finite density of Majorana particles. Existence of a first order phase transition and the associated TPS are the key prerequisites for the realization of the proposed mechanism. Within the TPS mechanism, the MZMs can be found throughout the system in contrast to the conventional mechanism that limits the existence of MZMs only to the edges. Furthermore, given that the on-site potential can be controlled via suitable application of gate voltages, the MZMs can be mobilized. Given that the implications of our study are general, it is not necessary to have exact realization of the toy model studied here in order to implement the mechanism in an experimental set-up. Nevertheless, for a proof-of-principle verification, the model is realizable in optically trapped ultracold atomic gases \cite{Bloch2008,Buhler2014,Zhang2015a}. Our results also provide a simple understanding of why the conventional idea of gap closing at topological transitions cannot be generally applicable to interacting systems. In general, our study establishes that the interplay of interactions and topology in many-particle quantum systems holds many, possibly useful, surprises.

{\it Acknowledgments:--}
We acknowledge the use of computing facility at IISER Mohali. We thank Ulrike Nitzsche for technical assistance. This project is funded by the German Research Foundation (DFG) via the projects A05 of the Collaborative Research Center SFB 1143 (projectid 247310070) and through the W\"urzburg-Dresden Cluster of Excellence on Complexity and Topology in Quantum Matter-ct.qmat (Project-ID 390858490-EXC 2147).
\bibliographystyle{apsrev4-2}
\bibliography{Shruti1,shreekant} 
\newpage

\onecolumngrid

\section{Mean-field BdG approach}
Starting with the Hamiltonian Eq. (1) in the main text, we decouple the interaction term using mean-field approach in density and pairing channels. The resulting Hamiltonian is given by,
\begin{eqnarray}
H = &-& \sum_i [\Tilde{\mu}_{i,A} ~c^{\dagger}_{i,A} c_{i,A}+ \Tilde{\mu}_{i,B} ~c^{\dagger}_{i,B} c_{i,B} ] \nonumber \\
&-& t \sum_i (c^\dagger_{i, A} c^{}_{i+1, A} + {\textrm H.c.}) \nonumber \\
&-& t' \sum_i (c^{\dagger}_{i, A} c^{}_{i, B} + c^{\dagger}_{i, B} c^{}_{i+1, A} + {\textrm H.c.}) \nonumber \\
&-& V' \sum_i (\Delta_{i,3} ~c^{}_{i+1,A} c^{}_{i,B} + \Delta_{i,2} ~c^{}_{i,B} c^{}_{i,A} + {\textrm H.c.}) \nonumber \\
&-& V \sum_i (\Delta_{i,1} ~c^{}_{i+1,A} c^{}_{i,A} + {\textrm H.c.}) + E_{{\rm const}},
\label{eqn:hmf}
\end{eqnarray}
where, 
\begin{eqnarray}
E_{{\rm const}} &=& \sum_i [ V |\Delta_{i,1}|^2 + V' (|\Delta_{i,2}|^2 +|\Delta_{i,3}|^2 )] \nonumber \\
    &+& \sum_i [V n_{i+1,A} n_{i,A} + V'  n_{i,B} (n_{i,A} +n_{i+1,A})], \nonumber \\
    \Tilde{\mu}_{i,A} &=& \mu + V (n_{i-1,A} + n_{i+1,A}) + V'( n_{i-1,B} + n_{i,B} ), \nonumber \\
    \Tilde{\mu}_{i,B} &=& \mu + V' (n_{i,A} + n_{i+1,A}). \nonumber 
\end{eqnarray}
Imposing spatial uniformity on local densities and the pairing amplitudes, we transform the Hamiltonian to Bloch basis using the standard Fourier transformations.
The mean-field Hamiltonian in Bloch space takes the form,
\begin{eqnarray}
H &=& \sum_{k} [\epsilon^{A}(k) c^{\dagger}_{k,A} c_{k,A} + \epsilon^{B}(k) c^{\dagger}_{k,B} c_{k,B}] \nonumber \\
&-& t' \sum_k [(1+e^{-i k}) c^{\dagger}_{k,A} c_{k,B} +{\textrm H.c.}]  \nonumber \\
&-& V \sum_k [\Tilde{\Delta}_{1}(k) ~c^{\dagger}_{k,A} c^{\dagger}_{-k,A} +{\textrm H.c.}] \nonumber \\
&-& V' \sum_k [\Tilde{\Delta}_{2}(k) c^{\dagger}_{k,A} c^{\dagger}_{-k,B} - \Tilde{\Delta}_{2}(-k)  c^{\dagger}_{k,B} c^{\dagger}_{-k,A} +{\textrm H.c.}] \nonumber \\
&+& E_{{\rm const}},
\end{eqnarray}
where,
\begin{eqnarray}
\Tilde{\Delta}_{1}(k) &=& 2 i \Delta_1 \sin k, \nonumber \\
\Tilde{\Delta}_{2}(k) &=& \Delta_2 - e^{i k} \Delta_3, \nonumber \\
\epsilon_A &=& -\mu - 2 n_A V - 2 n_B V' - 2t \cos{k}, \nonumber \\
\epsilon_B &=& - \mu - 2 n_A V', \nonumber \\
E_{{\rm const}} &=& N[V (n_{A}^2+ |\Delta_{1}|^2 ) + V' (2 n_{B}n_{A} + |\Delta_{2}|^2 +|\Delta_{3}|^2)]. \nonumber
\end{eqnarray}
It is convenient to represent the Hamiltonian in the following Nambu spinor notation, 
{\color{black}
\begin{eqnarray}
 H& = &\frac{1}{2}\sum_k \Psi_k^\dagger H_k \Psi_k,
\label{eq:hk0}
\end{eqnarray}
\noindent
with,
\begin{eqnarray}
\Psi_k &=&  \begin{bmatrix} c^\dagger_{k,A} & c^\dagger_{k,B}  & c_{-k,A} & c_{-k,B} \end{bmatrix}^\dagger,
H_k = \begin{bmatrix} \beta_k & \alpha_k \\ \alpha^\dagger_k & -\beta^T_{-k} \end{bmatrix}, \nonumber \\ 
\beta_k &=& \begin{bmatrix} -\Tilde{\mu}_A-2t\cos k & ~ ~ ~ ~ -t'(1+e^{-\mathrm{i}k}) \\ -t'(1+e^{\mathrm{i}k}) & -\Tilde{\mu}_B \end{bmatrix}, \nonumber \\
\alpha_k &=& \begin{bmatrix} -2\mathrm{i} V \Delta_1 \sin k & ~ ~ ~ ~ -V'\Delta_2(1-e^{-\mathrm{i} k}) \\ V'\Delta_2(1-e^{\mathrm{i} k}) & 0 \end{bmatrix}, 
\label{eq:hk}
\end{eqnarray}
\noindent
$\Tilde{\mu}_A = \mu+2n_AV+2n_BV'$ and  $\Tilde{\mu}_B = \mu + 2n_A V'$.
The Hamiltonian Eq. (\ref{eq:hk0}) acquires a diagonal form, $H = E_{0} + \sum_{k, \nu} E_{k,\nu} \gamma^{\dagger}_{k,\nu} \gamma^{}_{k,\nu}$, with the help of Bogoliubov transformations\cite{Zhu2016a},
\begin{eqnarray}
c^{}_{k,s} & = & \sum_{\nu} u^{\nu}_{k,s} ~\gamma_{k,\nu} + (v^{\nu}_{k,s})^{*} \gamma^{\dagger}_{-k,\nu} \nonumber \\
c^{\dagger}_{-k,s} & = & \sum_{\nu} v^{\nu}_{-k,s}\gamma^{}_{k,\nu} + (u^{\nu}_{-k,s})^{*} ~ \gamma^{\dagger}_{-k,\nu}
\end{eqnarray}
Local states for the Hamiltonian in real space may be denoted as
$E_0$ denotes the ground state energy and $E_{k,\nu}$ the excitation spectrum.} The constraint on the transformation to be unitary implies \(\sum_{\nu} |u^{\nu}_{k,s}|^2 + |v^{\nu}_{k,s}|^2 = 1\) and \(  u^{\nu}_{-k,s} = u^{\nu}_{k,s}\) \(v^{\nu}_{-k,s} = -v^{\nu}_{k,s}\)
\noindent
Notice that the basis of the Hamiltonian is doubled, making the solution redundant. Analysis have shown that \(E_{-k,\nu} = - E_{k,\nu}\)\cite{Zhu2016a}. Imposing this constraint, the mean-field parameters are computed in terms of the transformation coefficients as,
\begin{eqnarray}
n_{s}  &=& \frac{1}{N} \sum_{k, \nu} |u^{\nu}_{k,s}|^2~f(E_{k,\nu})
+|v^{\nu}_{k,s}|^2~(1-f(E_{k,\nu})) \nonumber \\
{\Delta_1} &=& - \frac{\mathrm{i}}{N} \sum_{k, \nu} \sin k~ u^{\nu}_{k,A} (v^{\nu}_{k,A})^{*} (2 f(E_{k,\nu}) -1) \nonumber \\
\Delta_2 &=& \frac{1}{N} \sum_{k, \nu} (u^{\nu}_{k,B} (v^{\nu}_{k,A})^{*} ~(1-f(E_{k,\nu})) - u^{\nu}_{k,A} (v^{\nu}_{k,B})^{*} ~f(E_{k,\nu}))
\end{eqnarray}
\noindent
{\color{black} where, $f$ denotes the Fermi function and $N$ the number of sites.
The mean-field parameters are determined self-consistently by starting with a random initial choice of the set \{$\Delta_1, \Delta_2, n_A, n_B$\}. We use $\delta_p = 10^{-4}$ and $\delta_n = 10^{-2}$ as the tolerance for the convergence condition on pairing and density parameters, respectively.

\section{Additional mean-field results}
\subsection{Phase Diagram}
\begin{figure}[H]
    \centering
    \includegraphics[width= 0.95\columnwidth,angle=0,clip=true]{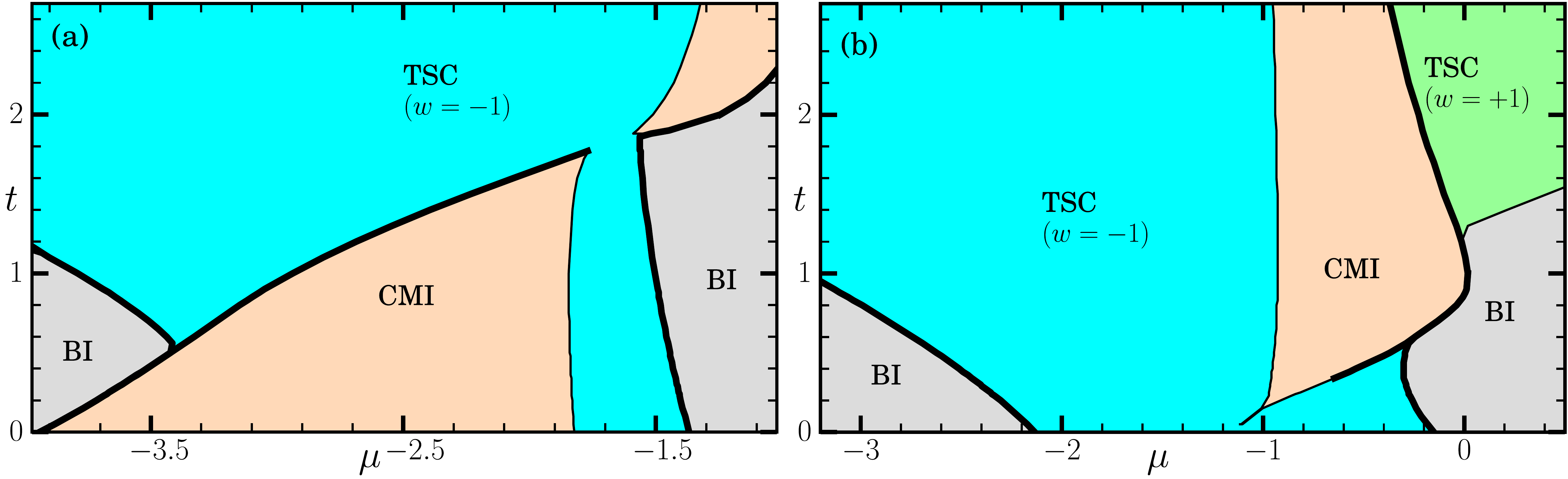}
    \caption{Ground state phase diagram in the $t-\mu$ plane for $V'=1$, and (a) $V=3$, (b) $V=0.3$}
    \label{fig:2}
\end{figure}
In order to check that the mean-field phase diagrams do not contain any qualitatively different phase from those already discussed in main text, we present the ground states for other representative choices of $V$ and $V'$. The phase diagram discussed in the main text corresponded to $V=V'=1$. Here we check for strong and weak $V/V'$ limits. We find that there is no qualitative change in the overall structure of the phase diagrams. Looking at the three phase diagrams together (see Figs. \ref{fig:2}(a) and \ref{fig:2}(b), and FIG. 2(e) in the main text), one can see the evolution of phases with changing $V'/V$. The $w=1$ TSC phase is missing in the $V > V'$ limit, and it occupies progressively larger area as $V'/V$ increases. Similarly, CMI phase dominates the $t-\mu$ phase diagram in the large $V'/V$ limit, and continues to appear in two disjoint pockets upon reducing $V'/V$. For $V'/V = 0.3$, one of the CMI regions disappears and the entire density range at low $t$ hosts TSC ($w=-1$) phase.

\subsection{Tunneling Density of States}
We investigate further the nature of different ground states in terms of sublattice-resolved tunneling density of states (TDOS), which is one of the physical quantities that is directly accessible via tunneling spectroscopy experiments. 
The sublattice-resolved TDOS is defined as, 
\begin{equation}
    N_s(\omega) = \frac{1}{N} \sum_{k,\nu} [|u^{\nu}_{k,s}|^2 \delta(\omega - E_{k,\nu})+|v^{\nu}_{k,s}|^2 \delta(\omega + E_{k,\nu})]
\end{equation}
where $N$ is number of sublattice sites, $u^{\nu}_{k,s}$ and $v^{\nu}_{k,s}$ are the electron-like and hole-like amplitudes of the Bogoliubov quasiparticles residing on sublattice $s$. For the calculations, the delta functions are approximated by the Lorentzians with broadening parameter $0.005$. Note that, except in the CMI phase, the TDOS spectra display two gaps. It is the gap at zero energy that is relevant as it stabilizes the order by lowering the total energy of the system compared to a non-gapped phase. The gap away from zero energy has its origin in the electronic bands in the non-interacting limit. The TDOS spectra also helps in differentiating between the SC gaps from those due to charge ordering. Note the asymmetry about zero in the sublattice-resolved TDOS for the case of CMI.
\begin{figure}[H]
    \centering
    \includegraphics[width =0.7\columnwidth]{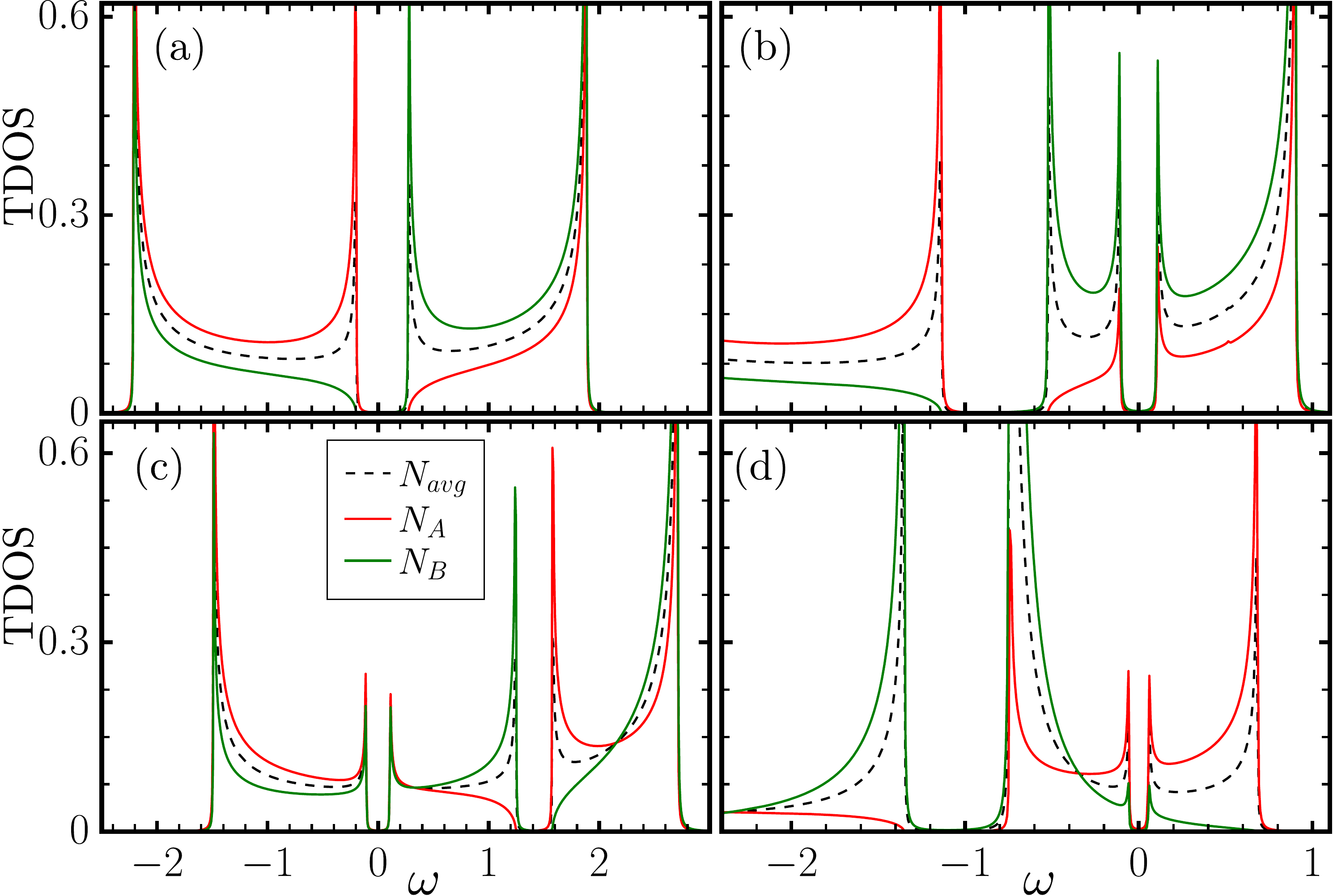}
    \caption{(a)-(d) Sublattice-resolved and average tunneling density of states (TDOS) for different ground states taken from the $V=1=V'$ phase diagram. (a) CMI: $t=0.1$, $\mu =-1.6$, (b) TSC with $w=-1$, $n>0.5$: $t=0.2$ , $\mu =-1.0$, (c) TSC with $w=-1$, $n<0.5$: $t=0.4$ , $\mu =-2.0$ and, (d) TSC with $w=1$: $t=2.0$ , $\mu =-0.2$.}
    \label{fig:tdos}
\end{figure}

Another very useful quantity that possesses spatially resolved information is the local DOS (LDOS). LDOS is defined as, 
\begin{equation}
    \textrm{LDOS} ~(i,E) =  \sum_s [|u^{\nu}_{i,s}|^2 \delta(E- E_{\nu})+|v^{\nu}_{i,s}|^2 \delta(E + E_{\nu})],
\end{equation}}
\noindent 
where, $u$ and $v$ are the elements of the real-space Bogoliubov transformation matrix.
LDOS at $E=0$ is especially relevant as it provides a first signature of the presence of Majorana modes in a superconductor. In the main text we have discussed LDOS at $E=0$ in the presence of quenched disorder (see FIG. 5 in  main text).

\noindent
\section{Local Entanglement Entropy}
\begin{figure}
    \centering
    \includegraphics[width= 0.65\linewidth,angle=0,clip=true]{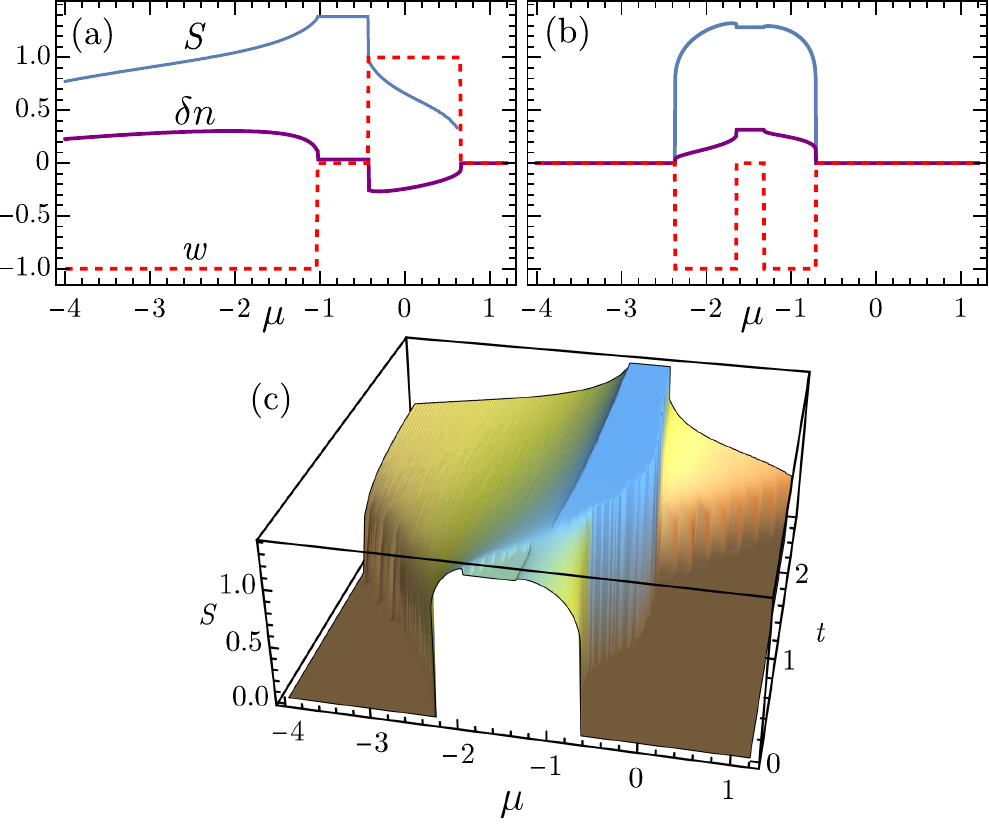}
    \caption{Entanglement entropy $S$, charge modulation $\delta n$ and winding number $w$ as a function of $\mu$ for, (a) $t = 2.3$ and (b) $t = 0.15$. (c) Surface plot of entanglement entropy in the $t-\mu$ plane. We have used $V=1=V'$ and $T=0$ for all the results shown in this figure.}
    \label{fig:EE}
\end{figure}
It has been proposed that entanglement entropy can be used efficiently to identify phase transitions \cite{Gu_2004}. We verify this proposal for the model under consideration by implementing the calculations of entanglement entropy. 
Local states for the Hamiltonian in real space may be denoted as
$\ket{00} = \ket{0}_{A} \ket{0}_B$, $\ket{10} = \ket{1}_{A}\ket{0}_B$, $\ket{01} = \ket{0}_{A}\ket{1}_B$, $\ket{11} = \ket{1}_{A}\ket{1}_B$. The presence (absence) of electron at site $i,A$ or $i,B$ is denoted by $1$($0$). The translational invariance of the Hamiltonian justifies the omission of index $i$ in the above notation. Local density matrix for any site is then given by tracing over all the other sites in the total density matrix.
\noindent
This is given by \cite{Gu_2004},
\begin{equation}
\rho_i =  z \ket{00}\bra{00} + a \ket{10}\bra{10} + b \ket{01}\bra{01} + d \ket{11}\bra{11}
	\label{eq:ldm}
\end{equation}
where,
\begin{eqnarray}
    d & = & \braket{\hat{n}_{i,A}~\hat{n}_{i,B}} \approx \braket{\hat{n}_{i,A}} \braket{\hat{n}_{i,B}}, \nonumber \\
    a & = & \braket{\hat{n}_{i,A}} - d, b = \braket{\hat{n}_{i,B}} - d, \nonumber \\
    z & = & 1 - a - b - d. 
	\label{eq:eeprm}
\end{eqnarray}
Note that all the parameters in Eq. (\ref{eq:eeprm}) are calculated within the mean field approximation. Von Neumann entropy, $S$, for the density matrix is then defined as,
\begin{eqnarray}
    S = - d\ln{d} - a\ln{a} - b\ln{b} - z\ln{z}.
	\label{eq:ee}
\end{eqnarray}
In order to test the usefulness of $S$ in determining the phase boundaries, we show the variation of $S$ with $\mu$ in the same plots that display the variations of $w$ and $\delta n$ (see FIG. \ref{fig:EE} (a)-(b)). We find that every instance of change in $w$ is accompanied by an anomaly in $S$. In order to test the usefulness of $S$ as an indicator of phase boundaries, we plot $S$($t$, $\mu$) as a surface plot (see FIG. \ref{fig:EE}(c)). Indeed, the entire phase diagram shown in FIG. 2(e) of the main text can be constructed based on this surface plot of entanglement entropy.

In the main text, we have mentioned that the sign of $w$ for topologically non-trivial phases are found to be perfectly correlated with the sign of $\delta n$. An interesting interpretation emerges if we consider the SO(3) theory of intertwined charge and superconducting orders \cite{Karmakar_2017}. Within this theory, charge order and superconductivity are combined into a vector order parameter with the $z$ component representing the charge order. Existence of a finite $\delta n$ fixes the reference for orientation of the loop that defines the winding number. Therefore the change in sign of $w$ may be interpreted as the change in the viewpoint from positive to negative $z$ axis, as controlled by the sign of $\delta n$.

\section{DMRG calculations}

\subsection{Averaged density distribution in the A and B chains}

\begin{figure}[tbh]
 \includegraphics[scale=0.8]{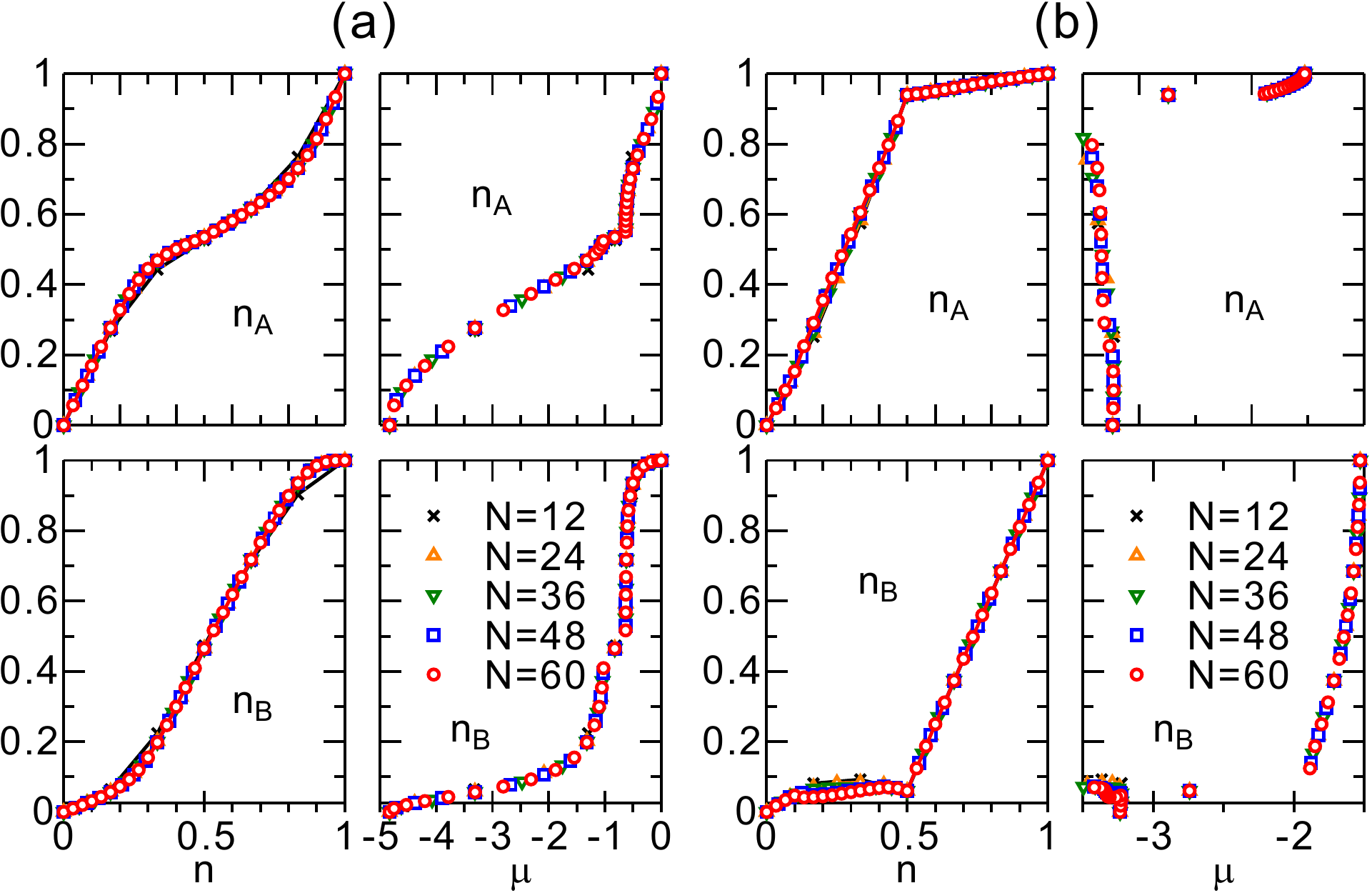}
 {
 \caption{Averaged density distribution $n_{\rm A}$ and $n_{\rm B}$ as
 	a function of filling $n$ as well as chemical potential $\mu$ for
 	(a) $t=2.0$, $t'=V=V'=1.0$ and (b) $t=t'=V'=1.0$, $V=3.0$.
 	Periodic chains are used.
 }
 }
 \label{fig:SM_density}
\end{figure}

It is useful to see how the spinless fermions are distributed over the A
and B sites. In Fig.~\ref{fig:SM_density} DMRG results for the densities
$n_{\rm A}$ and $n_{\rm B}$ are plotted as a function of filling $n$ as well
as chemical potential $\mu$. The chemical potential is estimated as
$\mu=-\partial E_0/\partial N_{\rm f}$, where $E_0$ is the ground-state
energy and $N_{\rm f}$ is total number of spinless fermions. We consider
two parameter sets: (a) $t=2.0$, $t'=V=V'=1.0$ and (b) $t=t'=V'=1.0$, $V=3.0$. In the case of (a) both $n_{\rm A}$ and $n_{\rm B}$ changes
continuously with increasing $n$ although $n_{\rm A}$ is larger than
$n_{\rm B}$ to gain the transfer energy. On the other hand, in the case
of (b) the most of fermions are distributed in the A chain at $n<0.5$
as a consequence of trivial phase separation (PS). Nevertheless, when the
densities are plotted as a function of $\mu$, we can see a jump at some
$\mu$, which corresponds to half filling $n=0.5$, for both of the
parameter sets. This is a `charge-modulated-insulators (CMI)
plateau' like a `Mott plateau' in the Hubbard model. A nearly-flat
behavior of $n_{\rm A}$ with $\mu$ is another typical feature of PS.

\newpage

\subsection{Phase separation}

\begin{figure}[tbh]
	\includegraphics[scale=0.6]{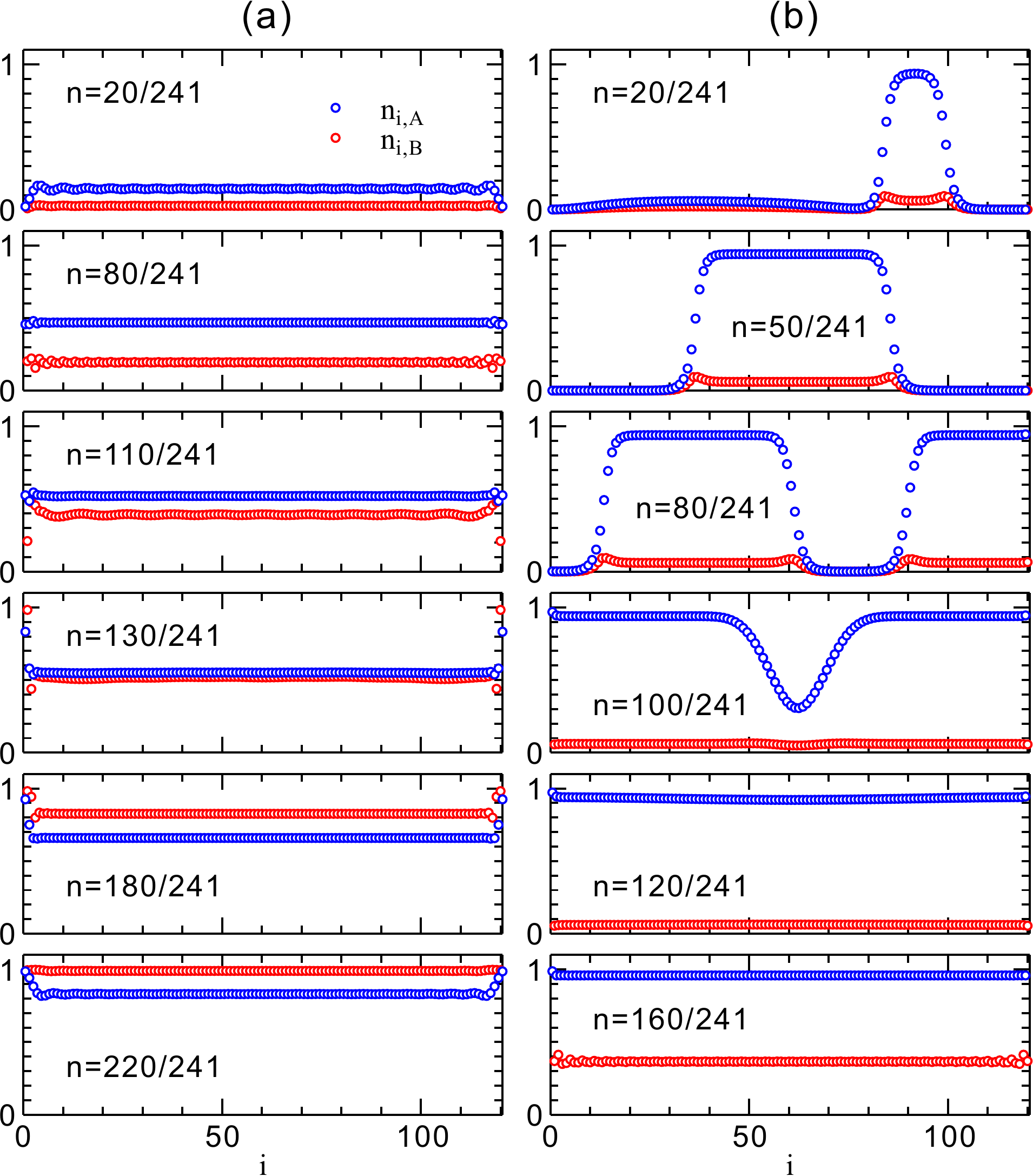}
	\caption{Local density distribution $n_{i,{\rm A}}$ and $n_{i,{\rm B}}$
		for (a) $t=2.0$, $t'=V=V'=1.0$ and (b) $t=t'=V'=1.0$, $V=3.0$.
	}
	\label{fig:SM_PS}
\end{figure}

For $t=t'=V'=1.0$, $V=3.0$ we found a trivial PS phase signified as
macroscopic condensate of fermions at low density. The PS can
be simply detected by looking at the local density distribution with
open boundary conditions in DMRG calculation. In Fig.~\ref{fig:SM_PS}
the local densities $n_{i,{\rm A}}$ and $n_{i,{\rm B}}$ are plotted
as a function of site $i$, where a $241$-site open chain is used. We can
clearly see a separation into particle-rich and particle-poor phases
at $n<0.5$ for $t=t'=V'=1.0$, $V=3.0$. Note that this PS is different
from the topological PS discussed in the main text. On the other hand,
a uniform distribution of the local densities for $t=2.0$, $t'=V=V'=1.0$
indicate the absence of trivial PS.

\newpage

\subsection{Density-density and Pair-Pair correlation functions}

\begin{figure}[tbh]
	\includegraphics[scale=0.8]{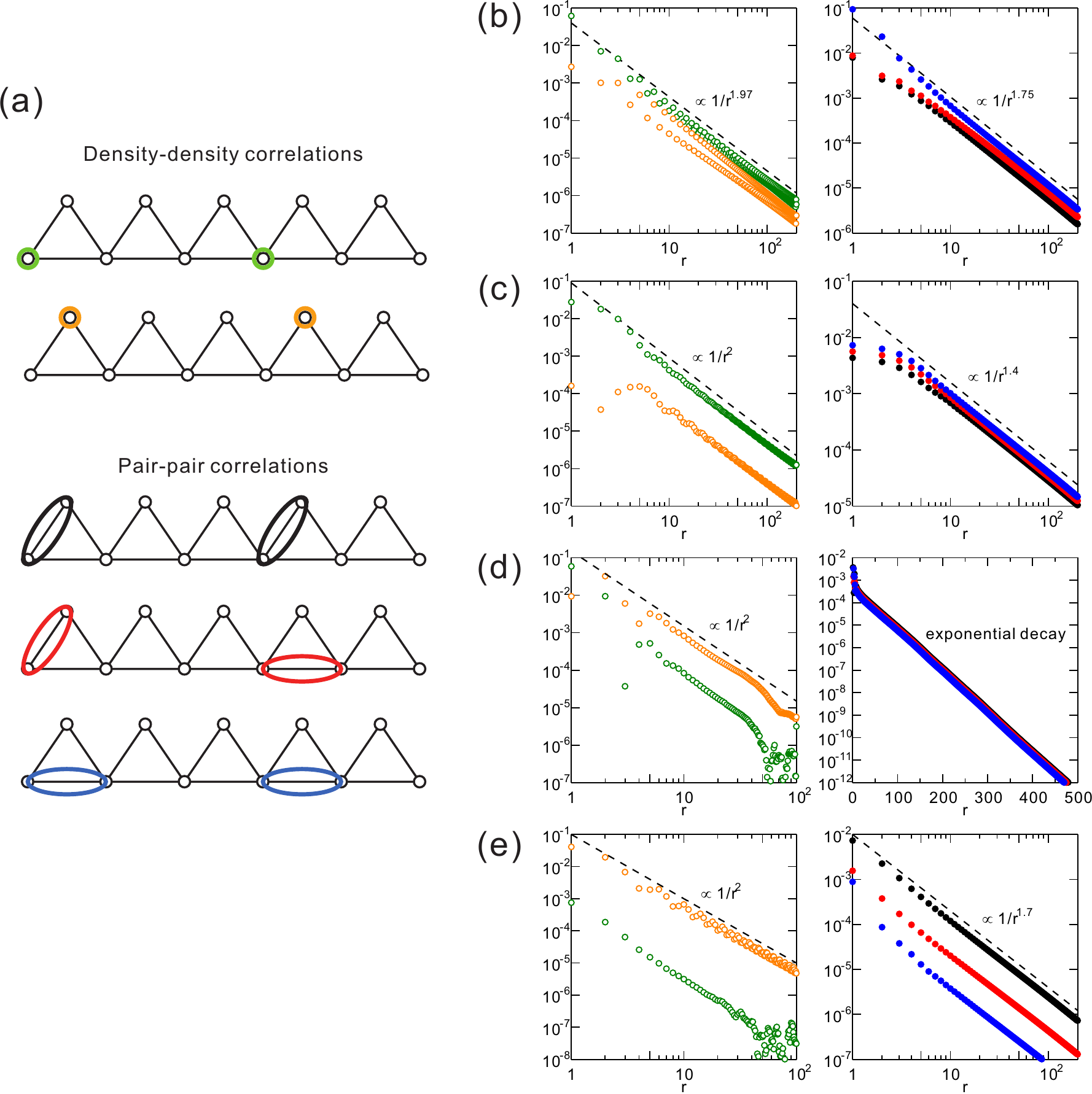}
	\caption{(a) Possible sets of sites and pairs for the density-density
		and pair-pair correlation functions, respectively.
		Density-density correlation functions (left panel) and SC
		correlation functions (right panel) for (b) $n=0.25$,
		(c) $n=0.6$, and (d) $n=0.9$ with $t=2.0$, $t'=V=V'=1.0$;
		(e) $n=0.625$ with $t=t'=V'=1.0$, $V=3.0$.
	}
	\label{fig:SM_corr}
\end{figure}

In 1D system the electronic phase is generally characterized by the
dominant correlation function with slower decay as a function of
distance. Depending on parameter and filling, our system has two
possible phases: charge density wave (CDW) and superconducting (SC).
Therefore, the density-density
$\langle\hat{n}_{i,s}\hat{n}_{j,s'}\rangle$ and pair-pair
$\langle P_{ii'ss'}^\dagger P_{jj'tt'}\rangle$, where
$P_{ijss'}=c_{i, s}^\dagger c_{j, s'}^\dagger$, correlation functions
are considered. As denoted in Fig.~\ref{fig:SM_corr}(a), we calculated
the density-density correlation function between two A sites (green)
or two B sites (orange); the SC correlation function between two pairs
of AB-AB (black), AA-AB (red), or AA-AA (blue). The DMRG results with
$N=2003$ open chain are shown in Fig.~\ref{fig:SM_corr}(b-e).
Both of the density-density and SC correlation functions exhibit
power-law decay with distance $r=|i-j|$ except $n\sim0.6$ for
$t=2.0$, $t'=V=V'=1.0$. Since the each correlation function shows
the same decay for any set of sites or pairs, 
the exponent $\gamma$ of
$\langle\hat{n}_{i,s}\hat{n}_{j,s'}\rangle \propto 1/r^\gamma$,
$\langle P_{ii'ss'}^\dagger P_{jj'tt'}\rangle \propto 1/r^\gamma$
can be uniquely estimated for a given parameter and filling.
The inverse exponent $1/\gamma$ is plotted as a function of filling
in Fig.~4(a,b) of the main text.

As mentioned above, the SC correlation function exhibits an exponential
decay instead of a power-law decay in the CDW phase
($ 0.43 \lesssim n \lesssim 0.75$) for $t=2.0$, $t'=V=V'=1.0$.
This means that our system deviates from the Tomonaga-Luttinger (TL)
liquid picture in this range. 

\newpage

\subsection{Tomonaga-Luttinger parameter}

\begin{figure}[tbh]
	\includegraphics[scale=0.8]{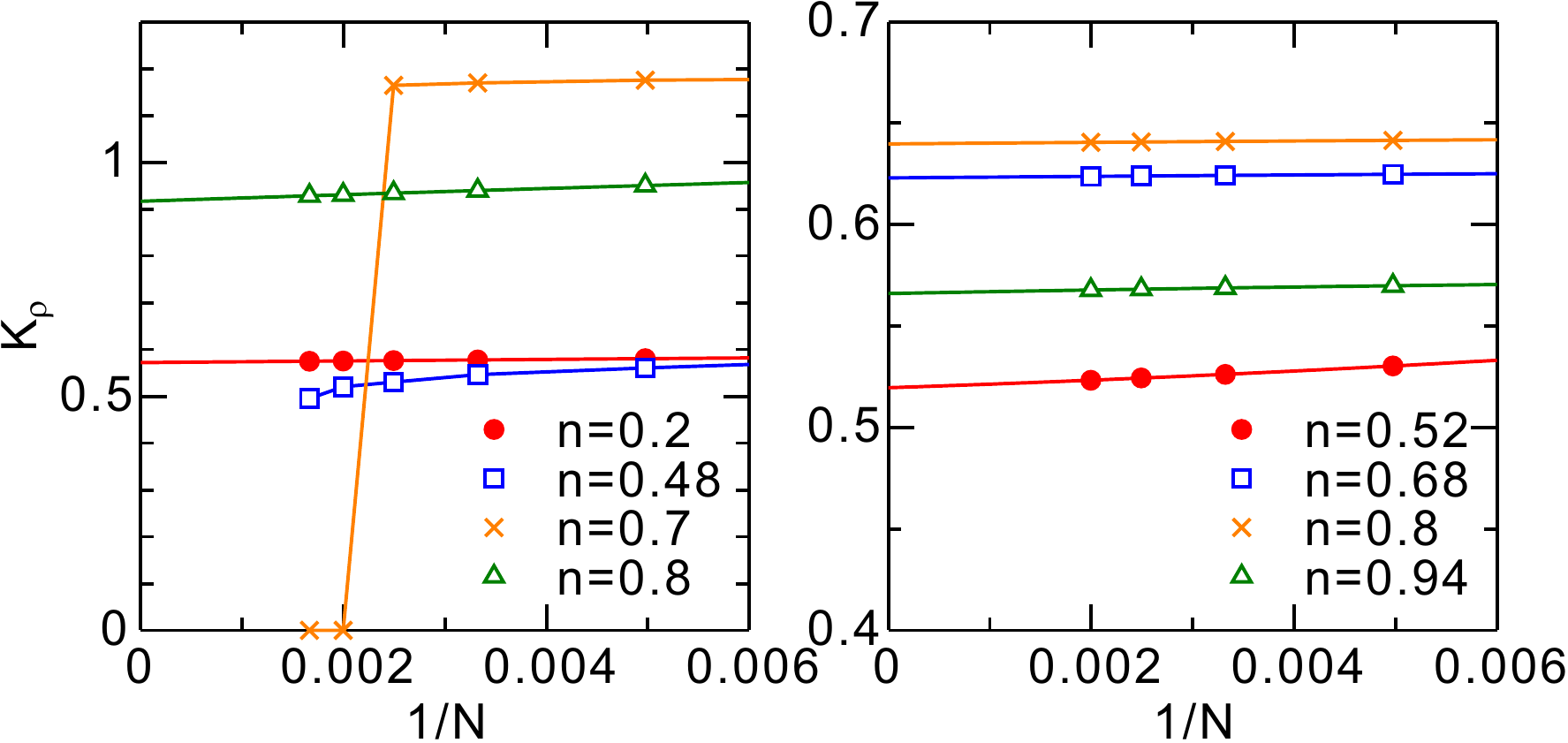}
	\caption{Finite-size scaling analysis of Tomonaga-Luttinger parameter
		for $t=2.0$, $t'=V=V'=1.0$ (left panel) and $t=t'=V'=1.0$, $V=3.0$
		(right panel). In the SC phase, fitting with 2nd polynomial
		function is performed.
	}
	\label{fig:SM_Krho}
\end{figure}

To confirm the validity of estimated exponent $\gamma$, we also calculated
the TL parameter $K_\rho$. It can be obtained accurately
via the derivative of charge structure factor at $q=0$ as
$K_\rho=\frac{1}{2} \lim_{q \to 0} \langle n(q) n(-q) \rangle$ with
$q=2\pi/N$ and $n(q)=\sum_l (e^{-iql}c_{l,A}^\dagger c_{l,A}+e^{-iq(l+1/2)}c_{l,B}^\dagger c_{l,B}$) using DMRG technique.
Since our system is spinless fermions, the noninteracting case corresponds
to $K_\rho=0.5$. The CDW and SC phases are identified by $K_\rho<0.5$ and
$K_\rho>0.5$, respectively. In Fig.~\ref{fig:SM_Krho} finite-size scaling
of $K_\rho$ is performed for several typical fillings. In the SC phase,
the scaling is simple and $K_\rho$ is easily extrapolated to a value
larger than $0.5$. On the other hand, in the non-TL range where the SC
correlation function exhibits an exponential decay with distance,
$K_\rho$ is extrapolated to $0$ in the thermodynamic limit.
The extrapolated values are plotted as a function of filling in
Fig.~4(a,b) of the main text. We can thus confirm that the relation
$K_\rho=1/\gamma$ is perfectly fulfilled at any filling.

\newpage

\subsection{Binding energy}

\begin{figure}[tbh]
	\includegraphics[scale=0.8]{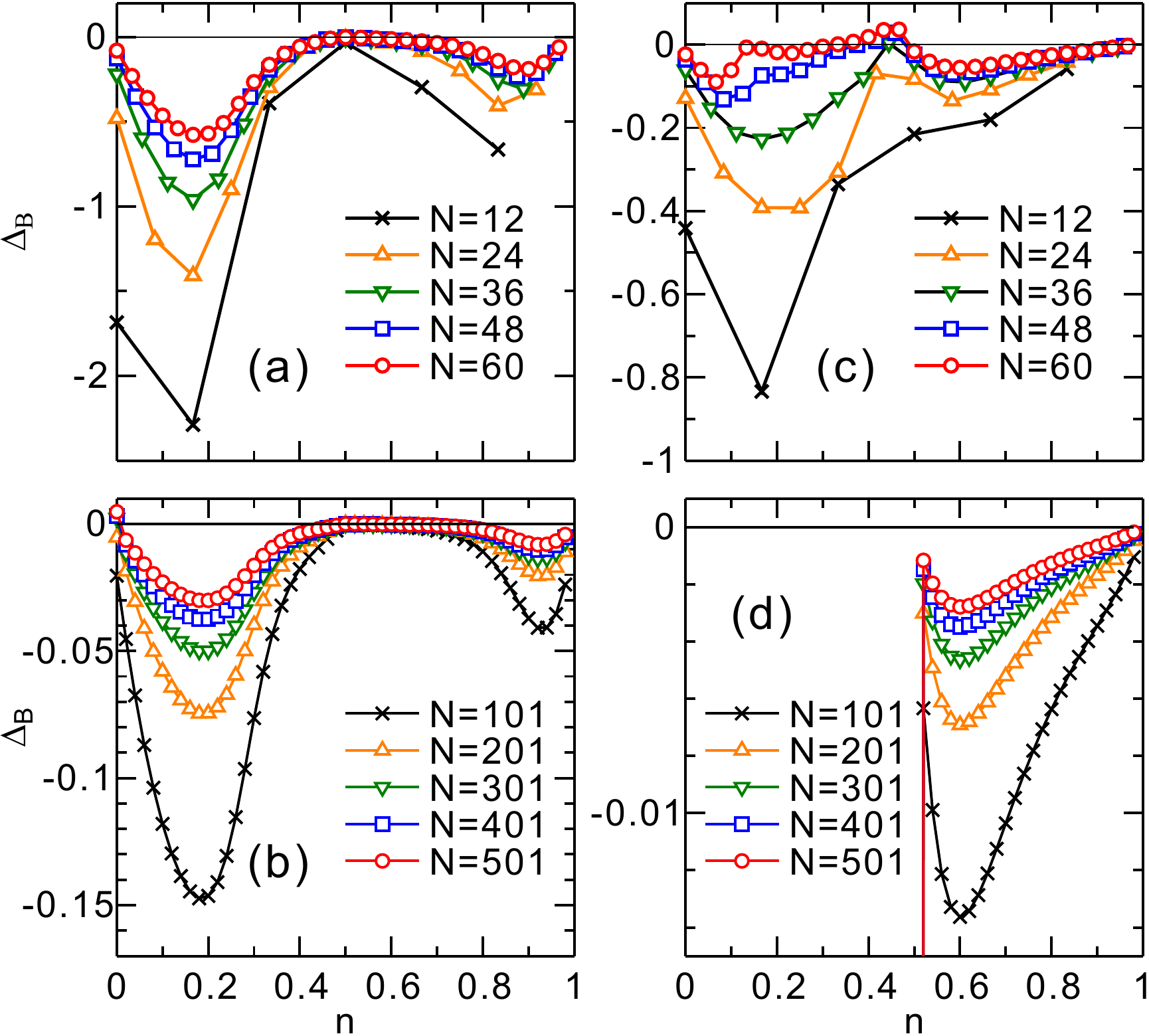}
	\caption{Binding energy as a function of filling for (a,b) 
		$t=2.0$, $t'=V=V'=1.0$ and (c,d) $t=t'=V'=1.0$, $V=3.0$.
		Periodic and open chains are used in (a,c) and (b,d), respectively.
	}
	\label{fig:SM_DeltaB}
\end{figure}

Finally, we calculated the binding energy between spinless fermions.
The binding energy is defined as
$\Delta_{\rm B}=[E_0(N_{\rm f}+2)-E_0(N_{\rm f})]-2[E_0(N_{\rm f}+1)-E_0(N_{\rm f})]$, where $E_0(N_{\rm f})$
is the ground-state energy of our system with $N_{\rm f}$ fermions. 
In Fig.~\ref{fig:SM_DeltaB} DMRG results for $\Delta_{\rm B}$ are
plotted as a function of filling. Both periodic and open chains are
used. Although finite $\Delta_{\rm B}$ with finite-size cluster may
indicate an enhancement of short-range pairing correlations,
it is extrapolated to zero in the whole range of $n$ and parameters except for the case of PS.

\end{document}